\documentclass[useAMS,usenatbib]{mn2e}
\usepackage{graphicx, epsfig, natbib}

\title[Modified gravity within the Solar System]{The influence of modified gravitational fields on motions of Keplerian objects at the far-edge of the Solar System}
\author[Michael Sokaliwska, Hans-J\"org Fahr, Pavel Kroupa]{Michael Sokaliwska$^{1}$\thanks{E-mail:; msokal@astro.uni-bonn.de}, Hans-J\"org Fahr$^{1}$\thanks{E-Mail:; hfahr@astro.uni.bonn.de} and Pavel Kroupa$^{1}$\thanks{E-mail:; pavel@astro.uni-bonn.de}\\
$^{1}$Argelander Institut f\"ur Astronomie, Auf dem H\"ugel 71, Bonn, Germany}

\begin{document}

\date{MNRAS, accepted}

\pagerange{\pageref{firstpage}--\pageref{lastpage}} \pubyear{2010}

\maketitle

\label{firstpage}

\begin{abstract}
We investigated the impact of three different modifications of Newtonian gravity on motions of Keplerian objects within the Solar System. 
These objects are located at distances of the order of the distance to the Oort cloud. With these three modifications we took into account a heliocentric Dark-Matter halo as was indicated by Diemand et al., Modified Newtonian Dynamics (MOND) and a vacuum-induced force due to a locally negative cosmological constant $\Lambda_-$ derived by Fahr and Siewert. In gravitationally bound systems it turns out that all three modifications deliver the same qualitative results: Initially circular orbits for the pure Newtonian case are forced to convert into ellipses with perihelion migrations. The quantitative consideration, however, of the orbital parameters showed strong differences between MOND on the one side, and Dark-Matter and $\Lambda_-$ effects on the other side.
\end{abstract}


\section{Introduction}

In modern astronomy there are some enigmas that physicists have not been able to solve yet. These enigmas seem to indicate the necessisty to depart from Newtonian gravity. One of these enigmatic phenomena is the so called Pioneer-Anomaly \citep{14} that seems to be an anomalous acceleration of the two space-craft, Pioneer 10 and 11, that were launched in the 1970's and move into two opposite directions of the Solar System. These spaceprobes show an acceleration towards the Sun of the order of $10^{-10}~\mathrm{m\,sec^{-1}}$ that seems to be independent of Solar distance, at least over Heliocentric distances between 20 and 70 AU, and cannot be explained by usual Newtonion dynamics. It seems that there exists an additional force responsible for this phenomenon, like modified gravity, but the constancy of the acceleration poses an important challenge.\\

\noindent\hspace{5mm} The second indication for modifed gravity manifests in the galactic rotation curve problem. The observed flat rotation curves at large galactic radii are not explainable by Newtonian dynamics alone.\\

\noindent\hspace{5mm}  This paper is inspired by these phenomena. We present in Section 2 of this paper three different modifications of gravity. Section 4 deals with the results of numerical calculations of the impact of these modifications and illustrates how these modifications act. The last section is a short outlook and suggests further possible investigations.

\section{Modifications of gravity}
Because Dark-Matter ($\Lambda$CDM) has become an essential ingredient of the standard scenario in modern cosmology the first modification that is investigated in the context of this paper is a heliocentric Dark-Matter halo as was indicated by \citet{16} from Neutralino Dark-Matter simulations.
\subsection{Heliocentric Dark-Matter minihalo}

In order to investigate the influence of a halo component of Dark-Matter it was necessary to choose a density-profile for the halo that was assumed to be bound to the Solar System and centred at the position of the Sun.\\

\noindent\hspace{5mm} According to \citet{14} who investigated whether a DM halo could be responsible for the Pioneer-anomaly we decided to use a halo that causes an additional acceleration towards the centre that is constant and independent on heliocentric distances, in view to indications given by the Pioneer-anomaly. We started with a halo-density-profile $\rho_{DM}\propto r^n$ where $n$ is a real number,
\begin{equation}
\rho_{DM}(r)=\rho_0r^n.
\end{equation}
Substituting this expression into the spherical Poisson-equation
\begin{equation}
 \frac{1}{r^2}\frac{d}{dr}\left(r^2\frac{d}{dr}\Phi_{DM}\right)=-4\pi G\rho_{DM},
\end{equation}
leads to the following expression for the gravitational force connected with the DM-halo
\begin{equation}
 F_{DM}=-4\pi G\frac{\rho_0r^{n+1}}{n+3}-\frac{C_1}{r}.
\end{equation}
If this force is constant over heliocentric distances, $n$ has to be chosen equal to $-1$ and the integration constant $C_1=0$. This then leads to the following density-profile
\begin{equation}
 \rho(r)=\frac{r_0\rho_0}{r},
\end{equation} 
where $r_0=1~\mathrm{AU}$ and $\rho_0$ is the density at $r_0$, that causes a constant additional acceleration towards the centre.

\subsubsection{Halo-mass}
The question is how the halo-mass can be selected such that the orbits of the outer planets, like Uranus, persist? For that purpose \citet{14} used data from the ephemeries of DE200 \citep{23} that contain Uranus' ephemeries down to an accuracy of $10^{-5}\mathrm{AU}$ in radial distance. The halo-mass enclosed within a given distance can be calculated via 
\begin{equation}
M_{DM}=4\pi\int_{r_0}^r r^2\rho_{DM}(r)dr.
\end{equation}
With that the halo-mass is fixed by the value of $\rho_0$. Calculating the deviation of the orbit of Uranus from the pure Keplerian case for different $\rho_0$ leads to the results shown in Fig. 1.

\begin{figure}[]
  \begin{center}
	\includegraphics[width=0.4\textwidth, angle=0]{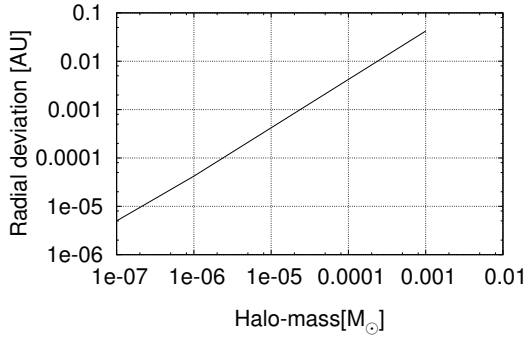}\caption{Deviation in radial distance $r$ for different Heliocentric DM 					halo-masses: One orbit of Uranus ($\approx 84~\mathrm{yr}$) has been integrated for both cases, for the usual Keplerian case and the case taking a $\frac{1}{r}$-Dark-Matter density profile into account. The differences in the absolute distance between the two cases has been calculated for every time-step. The enclosed mass is given for a radius of $50~\mathrm{AU}$}
\end{center}
\end{figure}

It is obvious that a radial distance deviation of the order of $10^{-5}~\mathrm{AU}$ is given for an enclosed halo-mass, within $50~\mathrm{AU}$,  $<10^{-6}~\mathrm{M_{\odot}}$ that corresponds to $\rho_0\approx 4.4\times10^{5}~\mathrm{M_{\odot}\,pc^{-3}}$. Such a halo-mass is in good agreement with the results from an investigation done by \citet{15,14}.\medskip

\noindent\hspace{5mm} Using this enclosed mass delivers a radial acceleration-component of the order of $10^{-12}~\mathrm{m\,sec^{-2}}$ which is two orders of magnitude smaller than the measured anomalous Pioneer-acceleration.\\

\textit{Parenthesis: Comparison with a standard-model}\\
\noindent\hspace{5mm} Is a Dark-Matter halo with a density-profile as given by eqn. (4) in good agreement with a general Navarro-Frenk-White (NFW) profile \citep{16,17,18}? The NFW-profile is given by\\

\begin{equation}
 \rho(r)=\frac{\rho_{0,NFW}}{\left(\frac{r}{r_s}\right)^{\gamma}\left[1+\left(\frac{r}{r_s}\right)^{\frac{1}{\alpha}}\right]^{(\beta-\gamma)\alpha}},
\end{equation}
where $r_s$ is the scale radius of the halo and $\rho_{0,NFW}$ is a constant density. For $\alpha=\gamma=1$ and $\beta=3$  this profile reduces to a standard NFW-profile
\begin{equation}
 \rho(r)=\frac{\rho_{0,NFW}}{\frac{r}{r_s}\left(1+\frac{r}{r_s}\right)^2}.
\end{equation}\smallskip

\noindent\hspace{0mm} In the limit of small distances compared to the scale radius ($r<<r_s$) this profile leads to the desired $\frac{1}{r}$-profile. Compared to Pioneer-like distances ($20-70~\mathrm{AU}$) $r_s$ has to be choosen very large, e.g. for $r=50~\mathrm{AU}$, $r_s > 5000~\mathrm{AU}$. With this the constant acceleration $a=2\pi Gr_s\rho_{0,NFW}\approx 10^{-12}~\mathrm{m\,sec^{-2}}$ leads to $\rho_{0,NFW}<47~\mathrm{M_{\odot}\,pc^{-3}}$.\\

\noindent\hspace{5mm} Using a correlation between realistic Halo-parameters (Darbringhausen et. al., in prepration), a halo with a scale radius $r_s>5000~\mathrm{AU}$ would have $\rho_{0,NFW}<30~\mathrm{M_{\odot}\,pc^{-3}}$. This value and the one stimated for a scale radius larger than $5000~\mathrm{AU}$ are of the same order of magnitude and it seems that the used density profile (4) is a realistic approximation.

\subsubsection{DM Critical Radius}
The distance from the centre where the Newtonian gravitional force due to the central mass (the Sun) equals the radial acceleration due to the halo with a density-profile according to (4) is called the critical radius. This radius turns out to be of the order of $r_{crit}\approx50000~\mathrm{AU}$, much larger than distances of about $70~\mathrm{AU}$, so the gravitational contribution of such a halo to the central gravity is negligible at such distances and would not lead to large daviations from pure Keplerian motion. \\

\medskip

\noindent\hspace{5mm} Objects expected to move at distances of the order of this critical radius are located in the Oort cloud. This cloud is assumed to be a reservoir of comets that represents the outer most material border of our Solar System and lies at a distance of about $50000~\mathrm{AU}$. \\

\noindent\hspace{5mm} So it is convenient to consider the influence of the DM-modification on Oort-cloud objects.The calculated additional radial acceleration was included into the used orbit integrating program that will be presented in Section 3.

\subsection{Modified Newtonian Dynamics (MOND)}

\subsubsection{Newtons second law}

\noindent\hspace{5mm} \citet{10} introduced a function $\mu(z)$ in Newtons second law where $z=\frac{a}{a_0}$, with the 'actual' acceleration $a$ devided by some critical acceleration $a_0$ which is a universal constant $\approx 2\times10^{-10}~\mathrm{m\,sec^{-2}}$,  derived from galactic rotation curves and turns out to be of the same order of magnitude as the Pioneer-acceleration. With this function Newton's second law is written 
\begin{equation}
 F=ma\mu\left(\frac{a}{a_0}\right)=mg_N,
\end{equation}
where $a$ is the actual acceleration of the object while $g_N=-\frac{GM_{\odot}}{r^2}$ is the strict Newtonian acceleration.\\

\noindent\hspace{5mm}The function $\mu(z)$ is not further specified but has the limiting-values

\begin{equation}
  \mu(z)  =  \left\{ \begin{array}{rcl}
                    1 & \mbox{if} & z\gg1, \\
                    z & \mbox{if} & z\ll1,
                    \end{array}\right.
\end{equation}
The second case corresponds to an actual acceleration $a$ that is much smaller than the critical acceleration $a_0$. This leads to an expression for $a$
\begin{equation}
 ma\mu\left(\frac{a}{a_0}\right)  \stackrel{a\ll{a_0}}{=}  m\frac{a^2}{a_0}=mg_N, \nonumber 
\end{equation}
such that

\begin{equation}
a  =  \sqrt{g_Na_0}.
\end{equation}

\noindent\hspace{5mm} This expression does not describe a continuous transition from the pure Newtonian regime ($a>>a_0$) to the pure MONDian regime ($a<<a_0$). So we chose a function that fullfills the required limit,
\begin{equation}
 \mu(z)=\frac{z}{1+z}.
\end{equation}
With this function equation (8) leads to 
\begin{equation}
a = \frac{g_N}{2}+\sqrt{\frac{g_N^2}{4}+g_Na_0},
\end{equation}
which is an expression for the actual acceleration that describes a continuous transition between the two regimes. This was the second type of acceleration considered for comparison studies in the orbit-integration routine.

\subsubsection{MOND-Critical radius}
In anology to the DM-case we calculated the radius were the gravitational acceleration due to the central mass equals the criticial (characteristic MONDian) acceleration $a_0$ to be $r_{crit}\approx 5000~\mathrm{AU}$. So also in the MONDian case there is no observable influence on Keplerian motion expected on scales of the planetary Solar System ($\approx 50~\mathrm{AU}$).

\subsection{Vacuum-type force: Similar to a negative cosmological constant}
The third modification of gravity considered here is a vacuum-induced force connected with a 'negative local cosmological constant' $\Lambda_-$.\\

\noindent\hspace{5mm} Our Universe is assumed to be describable by the Robertson-Walker-metric
\begin{equation}
ds^2=c^2dt^2-a^2(t)(dr^2+r^2(dr^2+sin^2\theta d\phi^2)),\\
\end{equation}
where $a$ is the scale parameter of the universe, $c$ the velocity of light, $r$ the distance and $dt$ an infinitisimal time-interval. $\theta$ and $\phi$ denote the polar and the azimuth-angles. This metric leads to an exact solution of the Einstein Field equations and describes a homogeneously matter-filled and isotropically curved universe expansion or contraction.\\

\noindent\hspace{5mm} However, gravitationally bound systems that have been decoupled from the homologous cosmic expansion due to gravitational collapse should show a different metrical dynamic behaviour than the residual universe.\\

\noindent\hspace{5mm} \citet{1} picked up this idea in order to describe the evolution of gravitionally bound systems like our Solar System, and started with an equation of motion given by
\begin{equation}
 \frac{d^2{x}}{dt^2}=-\frac{GM}{x^2}+\frac{\ddot{a(t)}}{a(t)}x,
\end{equation}
where $x$ describes the distance from the centre of the Solar System, $G$ is Newtons gravitational constant and $M$ the effective central gravitating mass. The first term on the RHS results from the attractive gravitational field whereas the second term corresponds to a repulsive force connected with the global expansion of the surrounding universe. This equation was derived by \citet{3} and describes the impact of the cosmic expansion on the dynamics within gravitationally bound systems (the 'local systems').\\

\noindent\hspace{5mm} Describing the extent of a local system by a "local scale factor" $l(t)$, that corresponds to the scale factor of a local RW-metric and considering the above equation at the border of this system leads to
\begin{equation}
 \ddot{l}(t)=-\frac{GM}{l^2(t)}+l(t)\frac{\ddot{a}(t)}{a(t)},
\end{equation}

\noindent\hspace{5mm}This second order differential equation can be transformed into a first-order one\footnote{The complete derivation of the following formalism can be reviewed in \citet{1}.} by multiplying with $2\dot{l}$ which leads to

\begin{equation}
\frac{d}{dt}\dot{l}^2(t)=2\frac{d}{dt}\frac{GM}{l}+\frac{\ddot{a}}{a}\frac{d}{dt}l^2.
\end{equation}

It can formally be integrated using a relation also derived by \citet[see Appendix there]{1} and leads to
\begin{equation}
 \dot{l}^2=2GM\left(l^{-1}-{l^{-1}_{rec}}\right)+\frac{1}{2}\left(\frac{\ddot{a}}{a}+\frac{\ddot{a}_{rec}}{a_{rec}}\right)\left(l^2-{l^2_{rec}}\right)+\dot{l}_{rec}^2.
\end{equation}

where $l_{rec}$ denotes the local scale factor at recombination-time $t_{rec}$ and $a_{rec}$ the global scale factor at this time.\\

\noindent\hspace{5mm} Solving this equation in its general form needs numerical methods and is not trivial. So it is convenient to analyze the asymptotic behaviour of this equation. 
At first consider the case that the scale factor $l$ at time $t$ is much larger than the associated scale factor at recombination time, $l(t)\gg l_{rec}$. Then the above equation reduces to
\begin{equation}
 \dot{l}^2=-\frac{2GM}{l_{rec}}+\frac{1}{2}\left(\frac{\ddot{a}}{a}+\frac{\ddot{a}_{rec}}{a_{rec}}\right)l^2+\dot{l}^2_{rec}.
\end{equation}
The RHS of this equation must always be positive and leads to the condition

\begin{equation}
 \frac{2GM}{l_{rec}}<\frac{1}{2}\left(\frac{\ddot{a}}{a}+\frac{\ddot{a}_{rec}}{a_{rec}}\right)l^2+\dot{l}^2_{rec}.
\end{equation}

\noindent\hspace{0mm}This condition then suggests that, if fullfilled, a region of "`gravitationally"' bound spacetime may under such conditions only grow large, if the central mass is sufficiently small. \\

\noindent\hspace*{5mm} In the opposite case, $l(t)\ll l_{rec}$, the equation of motion of the local scale factor yields
\begin{equation}
\dot{l}^2=\frac{2GM}{l}-\frac{1}{2}\left(\frac{\ddot{a}}{a}+\frac{\ddot{a}_{rec}}{a_{rec}}\right)l^2_{rec}+\dot{l}^2_{rec},
\end{equation}
which leads to the condition

\begin{equation}
\frac{2GM}{l}+\dot{l}^2_{rec}>\frac{1}{2}\left(\frac{\ddot{a}}{a}+\frac{\ddot{a}_{rec}}{a_{rec}}\right)l^2,
\end{equation}

\noindent\hspace{0mm}and can be interpreted as meaning that a sufficiently large central mass causes an asymptotically small vacuole of the Einstein-Straus type \citep{24}.\\

\noindent\hspace*{5mm}Using solutions of the well-known Friedmann-equations
\begin{equation}
a(t)  =  a_{rec}\left(\frac{t}{t_{rec}}\right)^{2/3},
\end{equation}
equation (18) leads to an equation
\begin{equation}
\left(\frac{\dot{l}}{l}\right)^2=\frac{\alpha}{l^3(t)}+\frac{\beta(t)}{l^2(t)}+\gamma(t),
\end{equation}
\noindent\hspace{0mm}with the parameter functions
\begin{eqnarray}
\alpha & = &2GM,\nonumber\\
\beta(t) & = & \left(\frac{2l_{rec}}{3t_{rec}}\right)^2-\frac{2GM}{l_{rec}}+\frac{l^2_{rec}}{9t^2_{rec}}\left(1+\frac{t^2_{rec}}{t^2}\right),\\
\gamma(t) & = & -\frac{1}{9t^2_{rec}}\left(1+\frac{t^2_{rec}}{t^2}\right).\nonumber
\end{eqnarray}

\noindent\hspace{0mm}that formally is very similar to the conventional Friedmann-equation that is given by

\begin{equation}
   \left(\frac{\dot{a}}{a}\right)^2=\frac{H_0^2\Omega_0}{a^3}-\frac{K}{a^2}+\frac{\Lambda c^2}{3},
\end{equation}

The comparison with the conventional Friedmann-equation shows a pronounced relationship between equation (24) and (26), despite some important differences between these two equations.\\

\noindent\hspace*{5mm}  The first term on the RHS of Eqn. (24) corresponds to the cosmological mass density term; the second term with $\beta(t)$ seems to correspond to the curvature term with the difference that this curvature is time-dependent here. The last term corresponds to a vacuum energy term, but is negative and time-dependent. This special result will be investigated.\\

\noindent\hspace*{5mm} Assuming the relation $\gamma(t)\approx \frac{\Lambda_-c^2}{3}$ leads to an equivalent of the "cosmological constant" that is today ($t=13,7\times10^9 ys$ and $t_{rec}=3\times10^5~\mathrm{yr}$) 

\begin{equation}
\Lambda_-\approx-1\times10^{-44}\frac{1}{m^2}. \nonumber                                                                                  
\end{equation}

As showed in \citet{6} the force that is acting on a particle, taking the influence of the vacuum into account, is given by
\begin{equation}
 \ddot{r}=-\frac{GM}{r^2}+\frac{c^2\Lambda}{3}r,
\end{equation}

\noindent\hspace{0mm}or in the case here,

\begin{equation}
 \ddot{r}=-\frac{GM}{r^2}+\frac{c^2\Lambda_-}{3}r,
\end{equation}

\noindent\hspace{0mm}where $\Lambda_-$ is the above derived negative constant which leads to an additional attractive force towards the centre. From this one can derive the corresponding potential via $-\frac{\partial \Phi(r)}{\partial r}=F(r)$,

\begin{equation}
 \Phi(r)=\frac{GM}{r}+\frac{c^2\Lambda_-}{6}r^2+const.
\end{equation}

The attraction derived from this potential increases with increasing distance $\propto{r^2}$. \citet{6} concluded that there is no observable influence of $\Lambda$ expected on scales of the inner Solar System. However, the above derived NEGATIVE cosmological constant $\Lambda_-$ is about three orders of magnitude larger than the conventional $\Lambda$ and thus changes earlier conclusions.

\subsubsection{$\Lambda_-$ Critical radius}
The force due to the negative vacuum energy density is $\frac{c^2\Lambda_-}{3}r$. Since $\Lambda_-$ is negative the induced force is directed towards the centre, just as the gravitational force.\\
\noindent\hspace*{5mm} The above expression leads to the condition for the critical radius,
\begin{equation}
 r_{crit}= \left(\frac{3GM}{c^2\Lambda_-}\right)^{\frac{1}{3}}.
\end{equation}
It turns out that this radius is about $r_{crit}\approx 4.5 \times 10^4~\mathrm{AU}$. At this distance from the Sun the influence of the vacuum cannot be neglected anymore, or, in other words, it is not possible to use perturbation theory because the "perturbation" is neither small nor time limited on a small interval. By comparison, the general cosmological constant with an upper limit of $10^{-47}~\mathrm{m^{-2}}$ would lead to a critical radius of about 8 light year.\\
\noindent\hspace*{5mm} Obviously a distance of the order of the critical radius is too large to expect a direct influence on planetary orbits. But if one takes the Oort cloud into account with an extent of about 1 light year, it should be possible to see deviations from pure Keplerian expectations. \\
\noindent\hspace{5mm}Since the critical radii of the three gravitational modifications that are investigated in this paper, are very large, they are not able to explain phenomena like the Pioneer-anomaly.

\section{The orbit-integrating program}
In order to investigate the impact of the three modifications of gravity derived above we use an orbit-integrator based on an embedded Runge-Kutta method of 4th order. The energy and angular momentum turns out to be a conserved quantity for all modifications.\\

\noindent\hspace*{5mm} Since the distances that were considered here are of the order the distance to the Oort cloud ($\approx 10^{4}$ AU), the whole problem reduced to a two-body problem: The central mass and the orbiting test-object.\\

\noindent\hspace*{5mm}We considered eleven cases with initial conditions that would cause clear circular orbits, -with initial distances in a range between $5000-~50000~\mathrm{AU}$- in a pure Keplerian potential and two cases that correspond to ellipses\footnote{These cases showed the same qualitative behaviour as the circular cases.}. 

The integration times were chosen differently (see Fig. 2 - 7) in a range between $4.9$ and $17.5~\mathrm{MYr}$.\\
\noindent\hspace*{5mm} It should be noted here that during the investigation of these three modifications every influence of external fields, like the Galactic tidal field, has been neglected. So only the gravitational force and the additional forces have been considered.

\section{Results of the orbit-integration}
Fig. 2-7 show the results of the orbit-integration for all three modifications. The solid line marks the circular orbit in a pure central-mass Keplerian case and the dashed line denotes the orbit under the respective gravitational modification.

\begin{figure}
  \begin{center}
	\includegraphics[width=0.4\textwidth, angle=0]{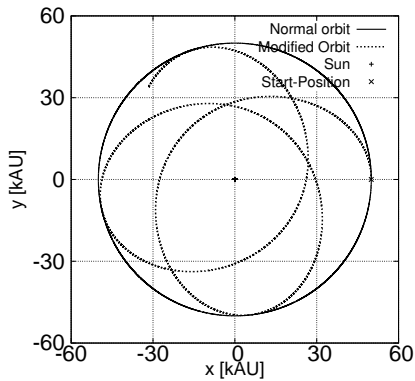}\caption{Resulting orbits for a distance of $50.000~\mathrm{AU}$ under the influence of a Dark-Matter minihalo. The integration time was $14.02~\mathrm{MYr}$.}
\end{center}
\end{figure}

\begin{figure}
  \begin{center}
	\includegraphics[width=0.4\textwidth, angle=0]{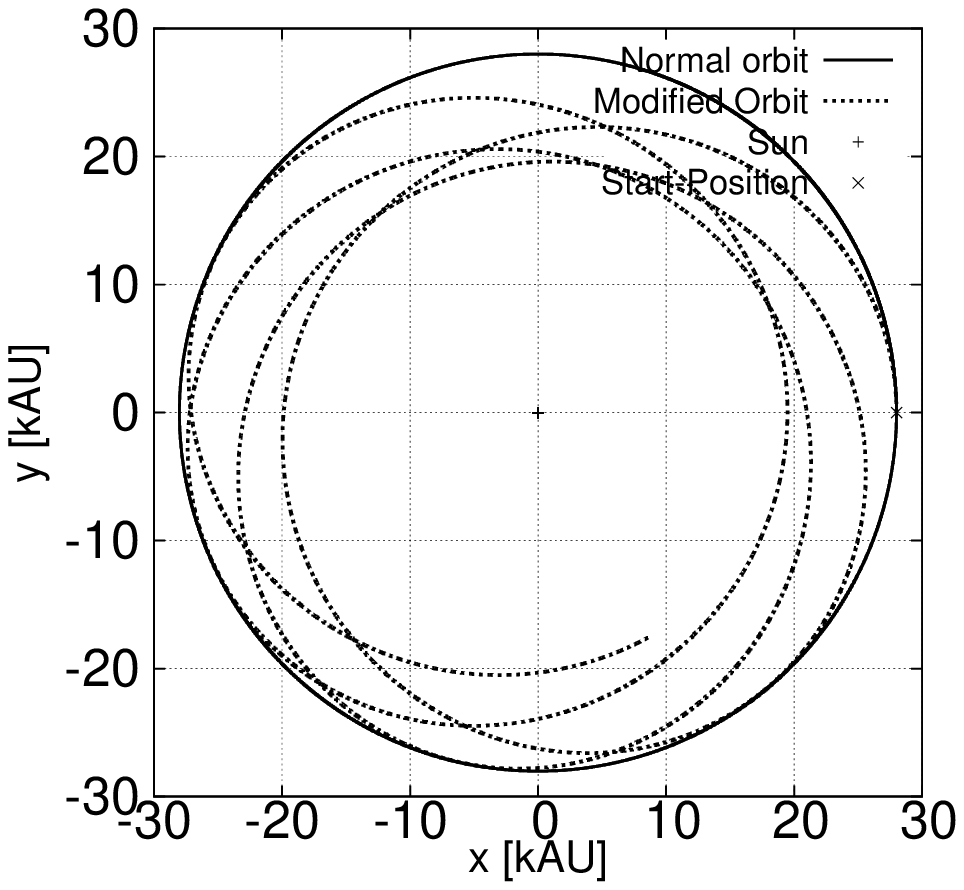}\caption{Resulting orbits for a distance of $28.000~\mathrm{AU}$ under the influence of a Dark-Matter minihalo. The integration time was $12.65~\mathrm{MYr}$.}
\end{center}
\end{figure}

\begin{figure}
  \begin{center}
	\includegraphics[width=0.4\textwidth, angle=0]{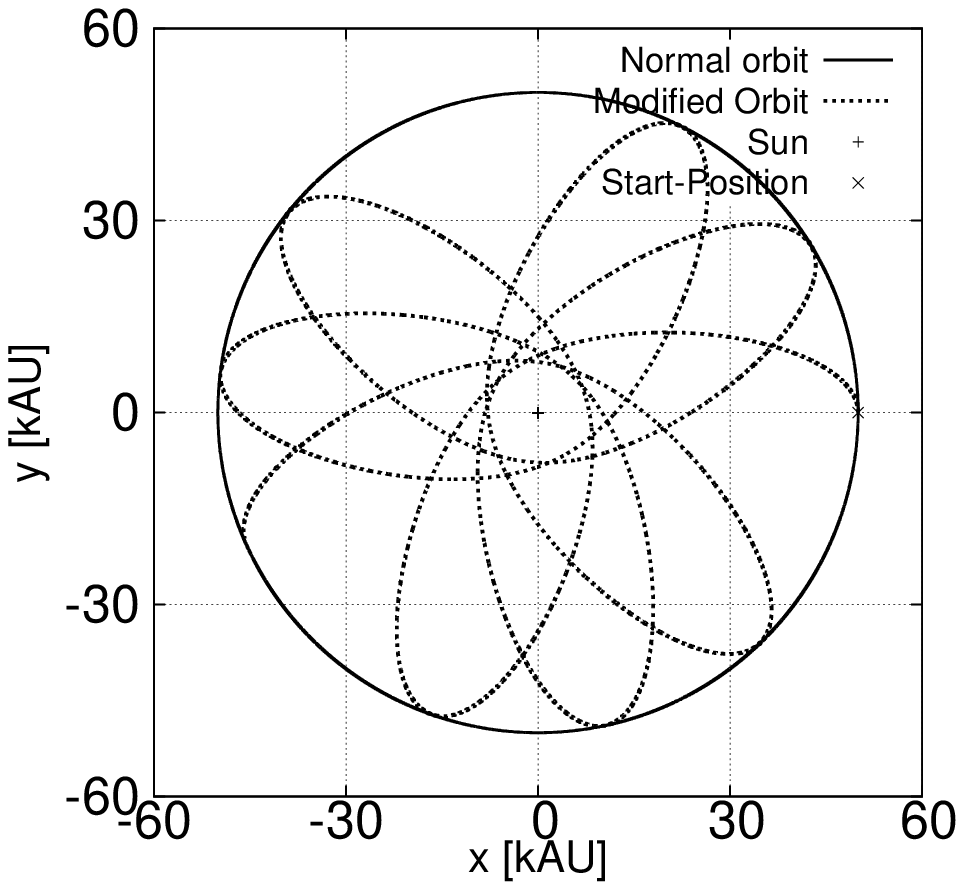}\caption{Resulting orbits for a distance of $50.000~\mathrm{AU}$ under the influence of MOND. The integration time was $12.3~\mathrm{MYr}$.}
\end{center}
\end{figure}

\begin{figure}
  \begin{center}
	\includegraphics[width=0.4\textwidth, angle=0]{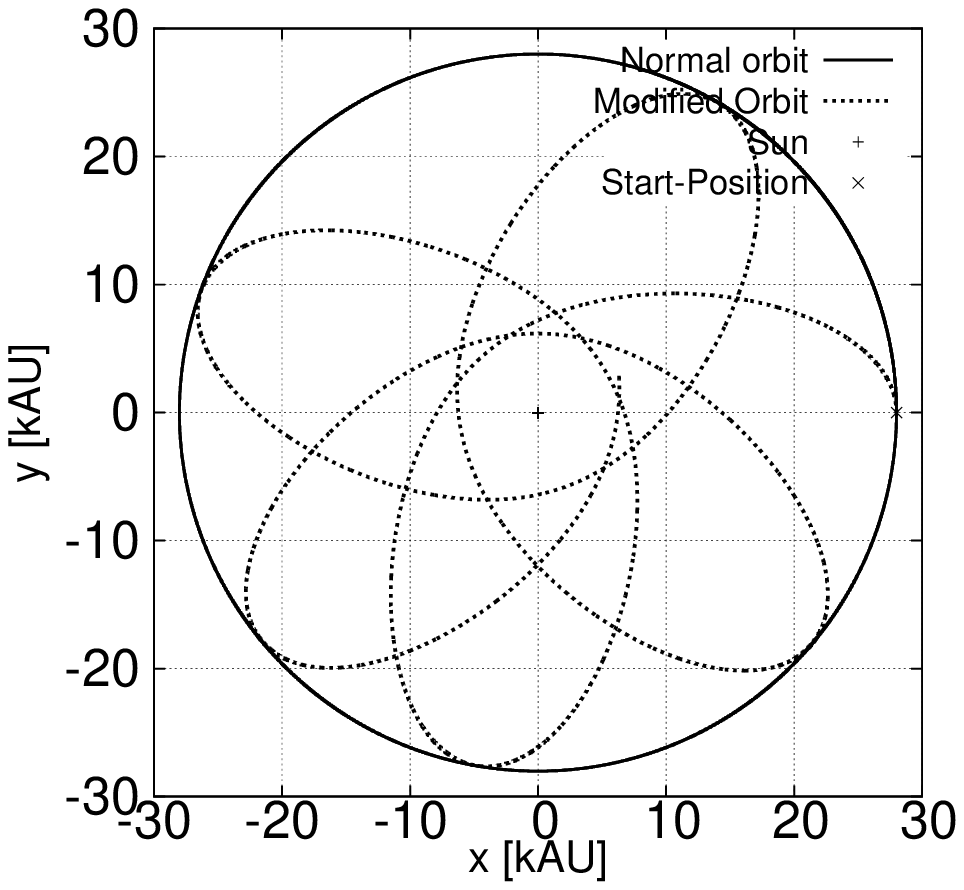}\caption{Resulting orbits for a distance of $28.000~\mathrm{AU}$ under the influence of MOND. The integration time was $4.9~\mathrm{MYr}$.}
\end{center}
\end{figure}

\begin{figure}
  \begin{center}
	\includegraphics[width=0.4\textwidth, angle=0]{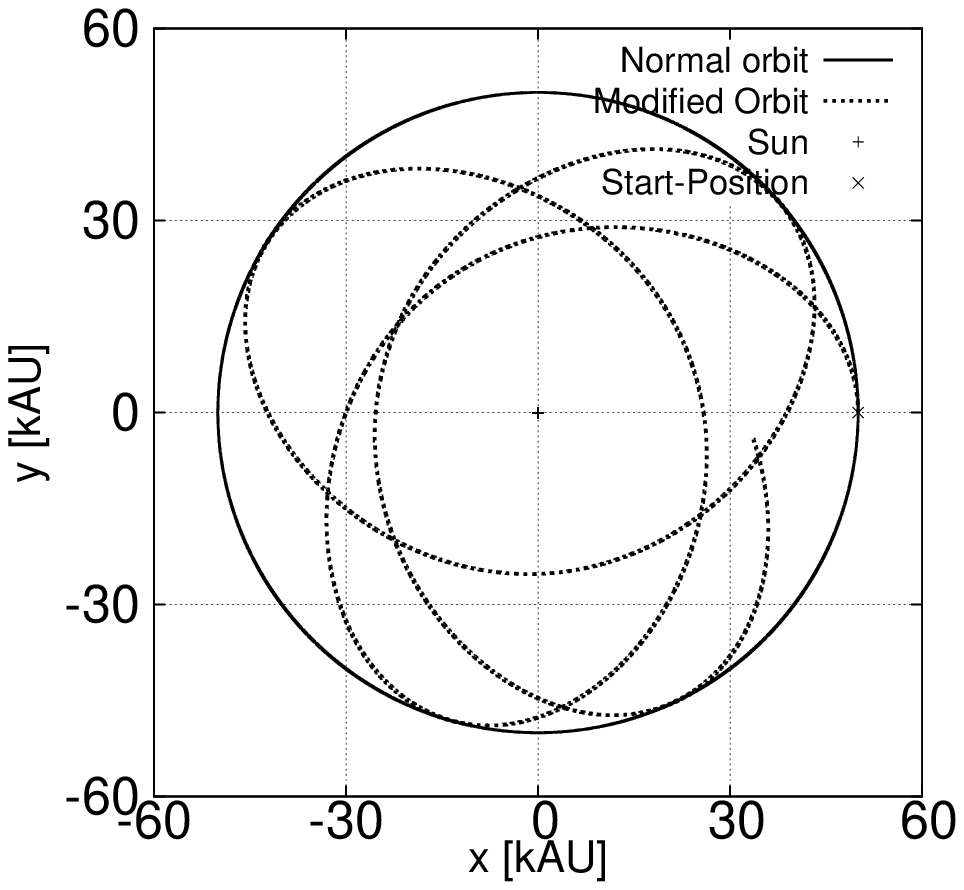}\caption{Resulting orbits for a distance of $50.000~\mathrm{AU}$ under the influence of $\Lambda_-$. The integration time was $17.5~\mathrm{MYr}$.}
\end{center}
\end{figure}

\begin{figure}
  \begin{center}
	\includegraphics[width=0.4\textwidth, angle=0]{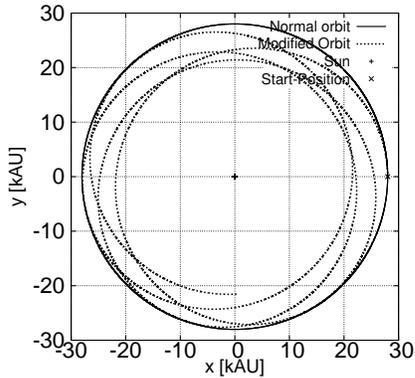}\caption{Resulting orbits for a distance of $28.000~\mathrm{AU}$ under the influence of $\Lambda_-$. The integration time was $13.6~\mathrm{MYr}$.}
\end{center}
\end{figure}

As one can see all three modifications cause strong deviations from pure Keplerian orbits around the central mass. Circular orbits become elliptical ones that in addition experience an aphelion-migration, a so called rosette-orbit. This qualitative behaviour occurs in all cases. The differences lie in parameters like the value of the aphelion-migration and the eccentricity for one turn. The results of the investigation of these parameters are presented in the next section.\\

\subsection{Orbital parameters}
In order to compare the impact of Dark-Matter, MOND and $\Lambda_-$ quantitatively we introduced two kinds of orbital parameters, the 'passive' orbital parameters and the 'active' ones.\\

\noindent\hspace*{5mm} The passive orbital parameters describe the orbits the test-object would be thought to move on under the assumption that no additional force or modification of gravity is valid. An observer who assumes a pure Keplerian potential would identify these parameters with the observed spatial and velocity components of the object. These parameters were calculated at every time step giving the true position and velocity values of the object. The passive eccentricity and the semi-major distance were the two passive orbital parameters considered here.\medskip

The second type of orbital parameters, the active ones, describe the actual shape of the orbit caused due to the gravitational modification. So the active eccentricity is the one of a fitted reference ellipse approximating one turn of the rosette-orbit, whereas the aphelion-migration describes how much the orbit turns per day.\\

\subsubsection{Passive orbital parameters}
The used orbit-integrating program was able to calculate instantaneously the passive orbital parameters for every time-step. The passive semi-major distance is given by

\begin{equation}
	a=\frac{r}{2-rv^2/\nu},
\end{equation}

\noindent\hspace*{0mm}where $a$ is then the passive semi-major-distance, $r$ the distance from the centre, $v$ the velocity and $\nu=GM_{\odot}$ Newtons gravitational constant times the central mass.\\
\noindent\hspace*{5mm} The passive eccentricity was then calculated via

\begin{equation}
\epsilon=\sqrt{1-\frac{p_{\phi}^2}{a\nu}},
\end{equation}

\noindent\hspace*{0mm}where $p_{\phi}$ denotes the constant angular momentum.\\
\noindent\hspace*{5mm} The calculation for the passive orbital parameters was done for one orbit with the initial conditions in velocity- and space-components corresponding to a circular Newtonian orbit with 50000 AU radius. Fig. 8-11 show the time-dependent behaviour of the semi-major-distance for all three modifications.\\
\noindent\hspace*{5mm} As one can see in all three modifications the semi-major-distance oscillates. In the case of Dark-Matter and $\Lambda_-$ the values of $a$ stay positive for all times. Not so in the MONDian case.

\begin{figure}
  \begin{center}
	\includegraphics[width=0.4\textwidth, angle=0]{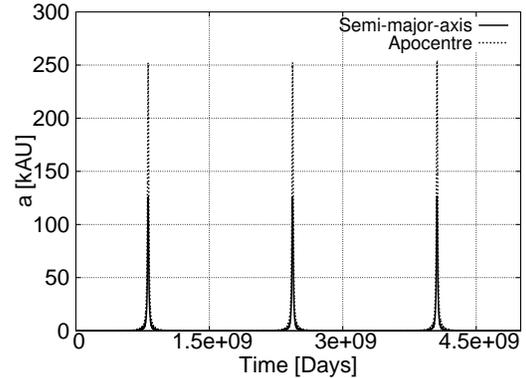}\caption{The behaviour of the instantaneous best-fitting (passive) semi-major-axis (SMA) and the aphelion distance with time under the influence of a Heliocentric Dark-Matter mini-halo. For the initial conditions a Newtonian circular orbit with $50000~\mathrm{AU}$ radius was used.}
\end{center}
\end{figure}

\begin{figure}
  \begin{center}
	\includegraphics[width=0.4\textwidth, angle=0]{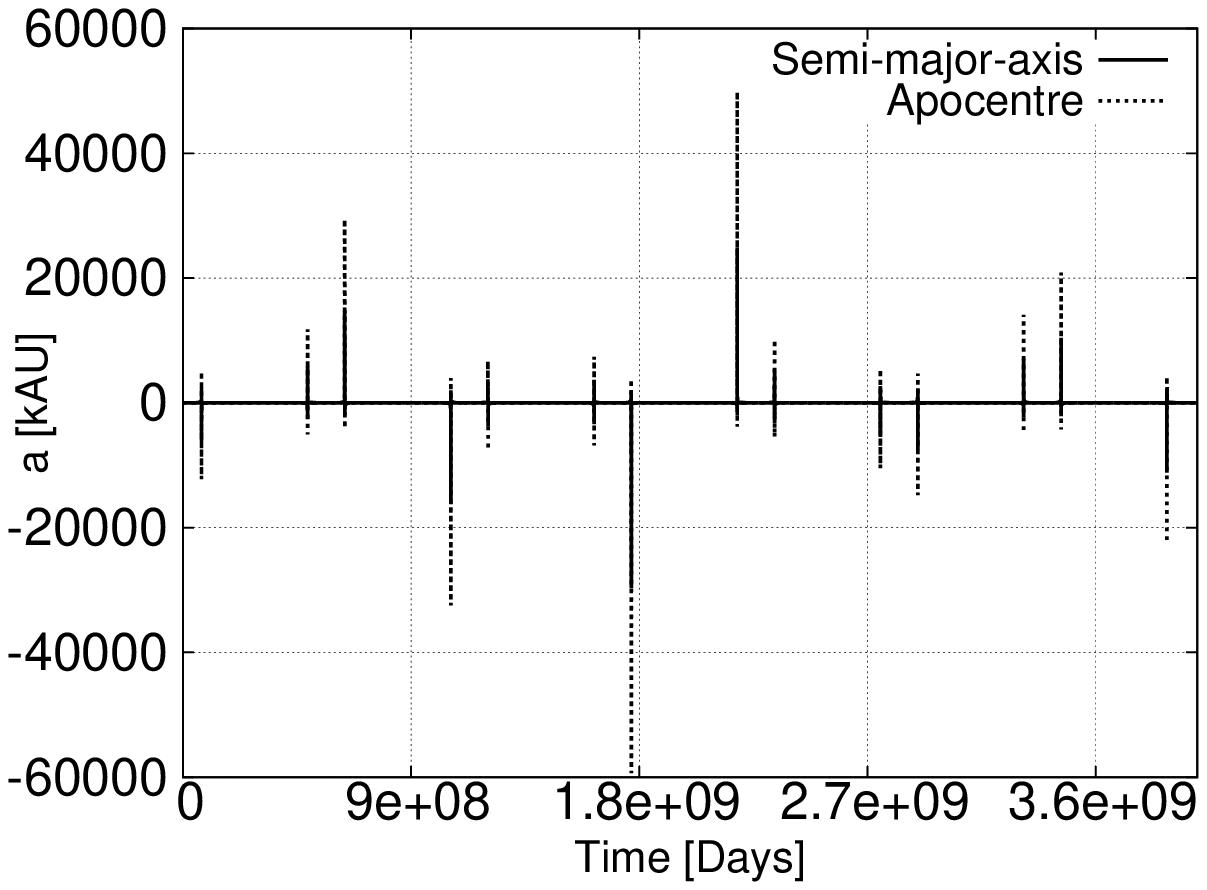}\caption{The behaviour of the instantaneous best-fitting (passive) semi-major-axis (SMA) and the aphelion distance with time under the influence of MOND. For the initial conditions a Newtonian circular orbit with $50000~\mathrm{AU}$ radius was used.}
\end{center}
\end{figure}

\begin{figure}
  \begin{center}
	\includegraphics[width=0.4\textwidth, angle=0]{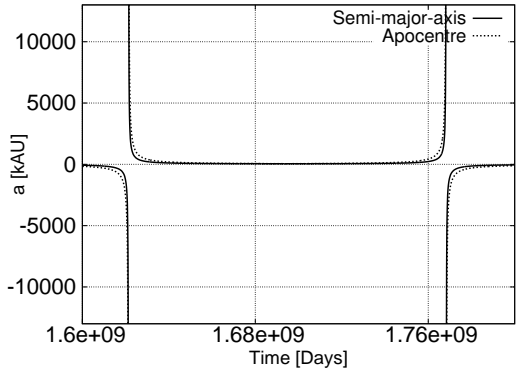}\caption{Closeup of the semi-major-axis (SMA) between two peaks (from Fig. 9). One can see clearly that $a$ takes positive and negative values, corresponding to closed and apparently hyperbolic orbits.}
\end{center}
\end{figure}

\begin{figure}
  \begin{center}
	\includegraphics[width=0.4\textwidth, angle=0]{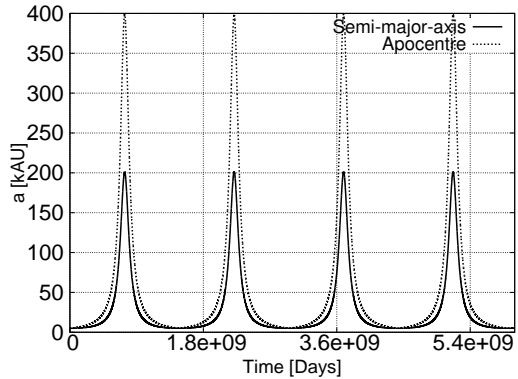}\caption{The behaviour of the instantaneous best-fitting (passive) semi-major-axis (SMA) and the aphelion distance with time under the influence of $\Lambda_-$. For the initial conditions a Newtonian circular orbit with $50000~\mathrm{AU}$ radius was used.}
\end{center}
\end{figure}

In MOND the semi-major-axis 'jumps' between positive and also negative values that correspond to open hyperbolic orbits. \\

\noindent\hspace*{5mm} So an observer who assumes a pure Keplerian potential would deduce such an object moving on a hyperbolic orbit around the Sun if he lives in a MONDian universe. Whereas Dark-Matter and $\Lambda_-$ would still produce closed orbits.\\

\noindent\hspace*{5mm} Here one can see one of the most important differences in this investigation between MOND on the one hand and $\Lambda_-$ and DM on the other hand.\medskip

Fig. 12-14 show the passive eccentricities for a circular orbit in pure Keplerian case under the impact of the three modifications.

\begin{figure}
  \begin{center}
	\includegraphics[width=0.4\textwidth, angle=0]{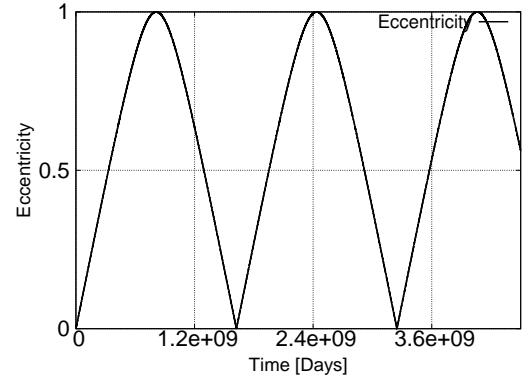}\caption{The behaviour of the instantaneous best-fitting (passive) eccentricity with time under the influence of a heliocentric DM-minihalo. For the initial conditions a Newtonian circular orbit with $50000~\mathrm{AU}$ radius was used.}
\end{center}
\end{figure}

\begin{figure}
  \begin{center}
	\includegraphics[width=0.4\textwidth, angle=0]{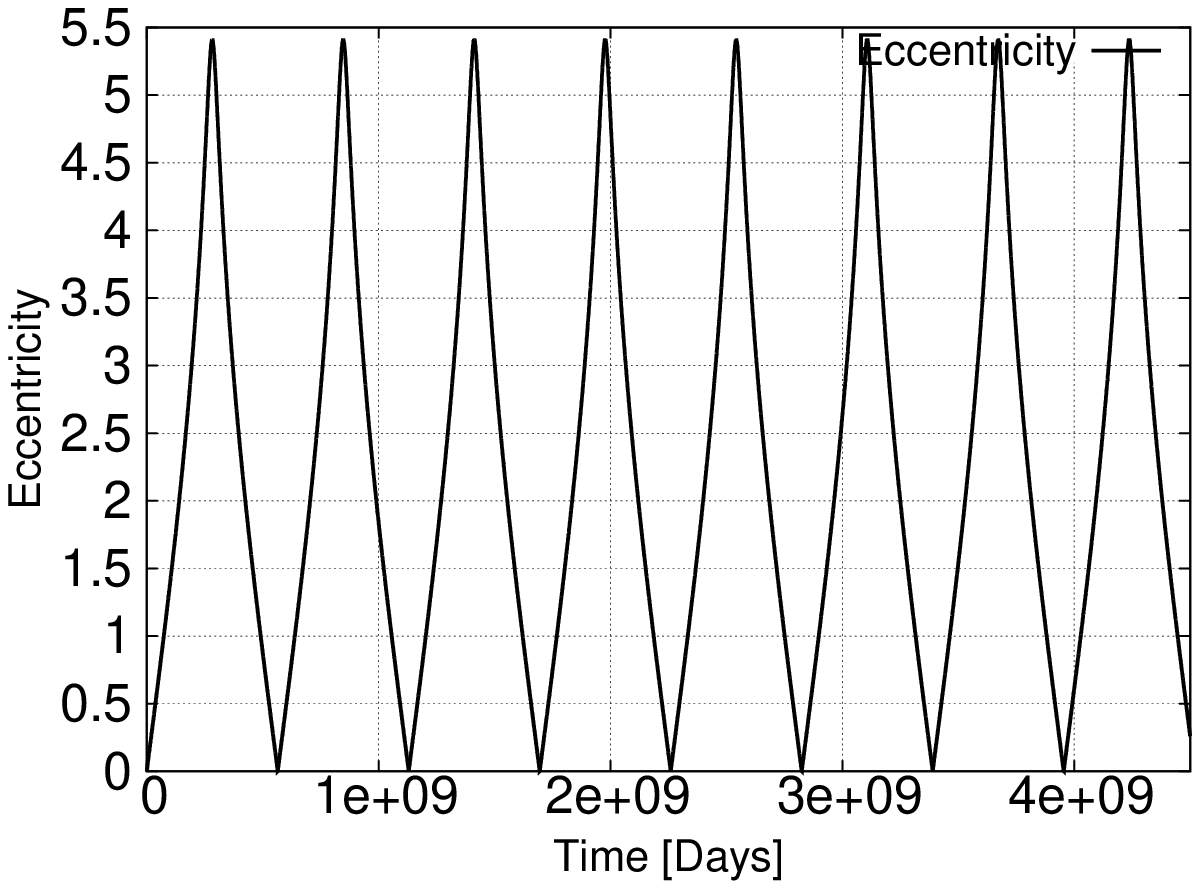}\caption{The behaviour of the instantaneous best-fitting (passive) eccentricity with time under the influence of MOND. For the initial conditions a Newtonian circular orbit with $50000~\mathrm{AU}$ radius was used.}
\end{center}
\end{figure}

\begin{figure}
  \begin{center}
	\includegraphics[width=0.4\textwidth, angle=0]{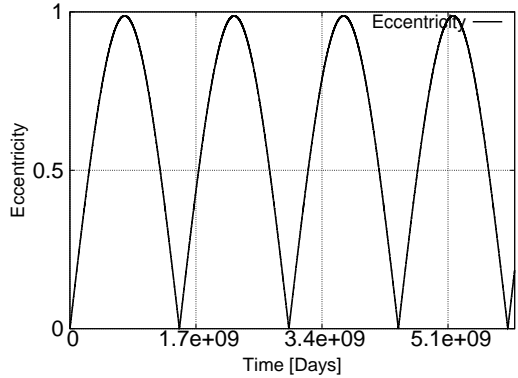}\caption{The behaviour of the instantaneous best-fitting (passive) eccentricity with time under the influence of $\Lambda_-$. For the initial conditions a Newtonian circular orbit with $50000~\mathrm{AU}$ radius was used.}
\end{center}
\end{figure}

The eccentricities in the DM and $\Lambda_-$ cases take values betwen $0$ and $1$ which corresponds to closed orbits. Not so in the MONDian case. Here the passive eccentricity has values in a range between $0$ and $5.5$ corresponding to the results for the semi-major-distance.

\subsubsection{Active orbital parameters}

The calculation of the active orbital parameters was done for every orbit with initial distances between $5000$ and $50000~\mathrm{AU}$. It should be mentioned here that the effective aphelion and perihelion of the turn of a rosette-orbit do not define the axis of symmetry of the turn of the orbit. Therefore we used a reference-ellipse approximating one turn of the orbit. So the following orbital parameters are with respect to this referenc-ellipse.\\

\noindent\hspace*{5mm} In order to calculate the active aphelion-migration of the generated rosette we took the coordinates of the aphelion of one turn and determined the angle of the aphelion with respect to the foregoing turn. The results for all three modifications are shown in Fig. 15 where we have fitted a curve to the data values which turned out to be highly dependent on the initial form of the orbit (e.g circular orbit, elliptical orbit, high or low eccentricities...)

\begin{figure}
  \begin{center}
	\includegraphics[width=0.4\textwidth, angle=0]{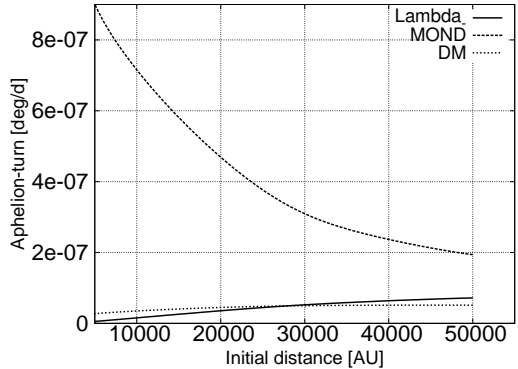}\caption{Comparison between the aphelion-migrations per day for all three modifications.}
\end{center}
\end{figure}

Very obvious is the coincidence of the curves of DM and $\Lambda_-$. MOND behaves completely differently. \\

\noindent\hspace*{5mm} A similar behaviour can be observed for the active eccentrities of one turn that are shown in Fig. 16: DM and $\Lambda_-$ seem to correspond very well. 

\begin{figure}
  \begin{center}
	\includegraphics[width=0.4\textwidth, angle=0]{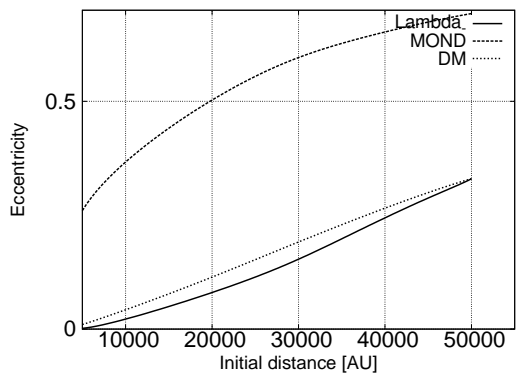}\caption{Comparison between the active eccentricities for all three modifications.}
\end{center}
\end{figure}

\section{Conclusions}
We investigated the influence of three modifications of gravity.\\
\noindent\hspace*{5mm}The first one was a Heliocentric Dark-Matter halo with a special density-profile, causing a distance-independent additional acceleration.\\
\noindent\hspace*{5mm} The second modification was in Newton's second law, $F=ma$, called MOND, whereas the third and last modification was a negative time-dependent cosmological ``constant", $\Lambda_-$, that causes an additional attraction towards the centre of the considered system.\\

\noindent\hspace*{5mm} The used Dark-Matter-halo had a special density profile $\sim\frac{1}{r}$ that generates a constant acceleration towards the centre, which was indicated by the Pioneer-acceleration.

\noindent\hspace*{5mm} All the modifications caused strong deviations from Keplerian circular orbits. Circular orbits are transformed into elliptical ones experiencing in addition an aphelion-migration (rosette-orbit).

The analysis of the passive and active orbital parameters has shown strong differences between MOND on the one side and $\Lambda_-$ and Dark-Matter on the other one. The MONDian effect seems to dominate even for small distances of about $5000~\mathrm{AU}$ whereas the other two modifications vanish at such distances. An explanation for this difference can be found in the non-linearity of MOND\footnote{MOND deals with the absolute values of the acceleration} and the fact that the contribution to the pure Newtonian acceleration is fairly larger than for $\Lambda_-$ or DM.\\

\noindent\hspace*{5mm} Although the Dark-Matter density-profile has been chosen such that it causes a constant acceleration like the Pioneer-acceleration neither this modification nor the two others investigated here can explain the Pioneer-anomaly. The distances where these modifications would have a measurable effect are far beyond Pioneer-like distances.\\

\noindent\hspace*{5mm} However, every considered modification of gravity would cause effects that should be observable. In the future it would appear useful to do further investigations.\\
\noindent\hspace*{5mm} Every modification of gravity has been considered at large distance of about $5000-~50000~\mathrm{AU}$ and seems to be negligible on scales of the inner Solar ($\approx 50AU$) System. A direct observation at distances of the order of the Oort cloud would be very difficult since every observation is limited by the possible resolution and sensitivity of the used telescope and furthermore by the observation time that would be necessary to observe deviations from the pure central mass Newtonian case at such large distances. However, if future-telescopes will be able to resolve e.g. Oort Cloud objects directly a comparison between the deduced orbital parameters with the passive (and active) parameters presented in this paper, would confirm or exclude the validity of the considered modifications of gravity.\\
\noindent\hspace*{5mm} Another possibility besides the direct observation of long periodic Oort cloud objects could be an experiment on an Earth-bound satellite placed at one of the Lagrangean points. These points are quasi-stable points within the framework of the Three-Body-Problem where the net gravitational force of the bodies is almost zero. It is for this reason why these points would be appropriate to test the modifications due to $\Lambda_-$ and a Dark-Matter halo. To test MOND these points would not be appropriate because the external Galactic field which has been neglected in this paper is much larger than $a_0$ over the whole Solar System and so would not lead to any MONDian effects over the Solar System. Investigating the additional forces acting on an object at the Lagrangean points and comparing them with the forces due to $\Lambda_-$ or DM would be a possible test of modified gravity. In addition also other modifications could be tested with such an experiment. Furthermore it could be useful to apply $\Lambda_-$ on larger scales like binaries, disk-galaxies or dwarf-spheroidal galaxies. Also an investigation of the contribution of $\Lambda_-$ to the perihelion-turn of Mercury seems to be a worthwhile application.\\

\noindent\hspace*{5mm}Whatever the investigation and development of modified gravitational theories will create, it seems to be obvious that the pure Newtonian Gravity due to central masses is not the whole story if the Pioneer-anomaly cannot be solved otherwise.\\
\noindent\hspace*{5mm}One ansatz could be the one postulated by \citet{20}. They postulated that the Pioneer-anomaly is caused by a distance dependence of Newtons gravitational constant $G(r)$ on the basis of the metric-skew-tensor gravity (MSTG) and the scalar-tensor-vector gravity (STVG). It seems that this ansatz could be succesfull: ''We have demonstrated that the STVG theory can explain the Pioneer anomalous acceleration data and still be consistent with the accurate equivalence principle [...]`` (\citet{20}, p.3435).

\section*{Acknowledgments}

We would like to thank Michael Marks and Jan Pflamm-Altenburg for providing the original code of the orbit-integrating program.

\bibliographystyle{mn2e}
\bibliography{biblio.bib}
\makeatletter   \renewcommand{\@biblabel}[1]{[#1]}   \makeatother 

\label{lastpage}

\end{document}